\documentclass[twoside]{dis08}
\usepackage[latin1]{inputenc}
\usepackage[dvips]{graphicx,epsfig,color}
\usepackage{wrapfig,rotating}
\usepackage{amssymb,amsmath,array}

\begin{document}
\newcommand{\be}{\begin{equation}} \newcommand{\ee}{\end{equation}}
\newcommand{\ba}{\begin{eqnarray}} \newcommand{\ea}{\end{eqnarray}}
\newcommand{\bea}{\begin{eqnarray}} \newcommand{\eea}{\end{eqnarray}}
\newcommand{\bean}{\begin{eqnarray*}} \newcommand{\eean}{\end{eqnarray*}}
\newcommand{\bm}[1]{\mbox{\boldmath $#1$}}
\newcommand{\s}[1]{{\scriptscriptstyle #1}}
\newcommand{\st}{{\s T}}
\def\slash{\rlap{/}}
\title{Single spin asymmetries and gluonic pole matrix elements}

\author{P.J.\ Mulders
%
%
\vspace{.3cm}\\
%
VU University - Department of Physics and Astronomy\\
De Boelelaan 1081 - 1081 HV Amsterdam - Netherlands
%
}

\maketitle

\begin{abstract}
We investigate the emergence of single spin asymmetries (SSA) in hard
processes using transverse momentum dependent (TMD) distribution and 
fragmentation functions.  Specifically, the description of SSA involves 
time reversal-odd functions.  Process-dependence (non-universality) in 
measurements of SSA can be attributed to the non-trivial gauge link 
structure in the TMD correlator. Finding the appropriate gauge links,
however, also enables us to characterize the 
non-universality~\cite{url,Bomhof:2007xt}. 
\end{abstract}
\section{Introduction}
In recent years many theoretical and experimental studies aimed for an
understanding of the mechanisms that lead to single spin asymmetries (SSA)
in hard hadronic scattering processes. In collinear approximation
(integrating over all transverse momenta) all leading twist distribution
(and fragmentation functions) only depend on the longitudinal momentum
fraction $x$ (or $z$) and involve double spin asymmetries, i.e.\ polarized
quarks are only found in polarized hadrons (and vice versa). 
Single spin asymmetries (SSA) involve twist-three collinear 
quark-gluon matrix elements. In the specific
limit of a zero-momentum gluon, referred to as {\em gluonic pole
matrix elements} such as the Qiu-Sterman matrix 
elements~\cite{Qiu:1991pp},
the effects can appear at leading order.
Also in model calculations the effects of these soft gluon interactions
between the target remnant and the hard part parton have been
demonstrated, giving rise to specific effects for initial or final state 
interactions~\cite{Brodsky:2002cx}.

Going beyond the collinear approximation and including the effects of 
intrinsic transverse momenta of partons provides another mechanism to 
generate leading order SSA, which can be traced back to correlations 
between the intrinsic transverse motion and spin of partons and/or hadron.
The effects are described by transverse momentum dependent 
(TMD) distribution functions, containing both T-even and T-odd parts
and depending on longitudinal momentum fraction
$x$ and the transverse momentum $p_{T}$ as appearing in the Sudakov
decomposition $p = x\,P + p_{T}$ (or $p = (1/z)\,P + p_{T}$ for 
fragmentation). The TMD correlators include Wilson lines, which besides
ensuring gauge-invariance are in the case of distribution functions the
sole cause of T-odd contributions. Upon $p_{T}$-integration one finds
after weighing with $p_{T}$ the socalled {\em transverse moments}
of the TMD distribution functions, which can be separated into T-even
and T-odd parts that are universal and of which the T-odd part can be
identified with the gluonic pole matrix elements. 

\section{\label{TMDcorrelators}
Transverse momentum dependent (TMD) correlators}

The TMD distribution functions are projections 
of the TMD quark correlator defined on the light-front
(LF: $\xi{\cdot}n\,{\equiv}\,0$)
\begin{equation}\label{TMDcorrelator}
\Phi_{ij}^{[C]}(x{,}p_\st{;}n)
={\int}\frac{d(\xi{\cdot}P)d^2\xi_\st}{(2\pi)^3}\ e^{ip\cdot\xi}\,
\langle P{,}S|\,\overline\psi_j(0)\,\mathcal U_{[0;\xi]}^{[C]}\,
\psi_i(\xi)\,|P{,}S\rangle\big\rfloor_{\text{LF}}\ .
\end{equation}
The \emph{Wilson line} or \emph{gauge link}
$\mathcal U_{[\eta;\xi]}^{[C]}\,
{=}\,\mathcal P{\exp}\big[{-}ig{\int_C}\,ds{\cdot}A^a(s)\,t^a\,\big]$
is a path-ordered exponential along the integration path $C$ with
endpoints at $\eta$ and $\xi$, ensuring gauge-invariance.
In the TMD correlator~\eqref{TMDcorrelator} the integration path $C$ in
the gauge link is process-dependent. 

\begin{wrapfigure}{l}{0.5\columnwidth}
\includegraphics[width=3.0cm]{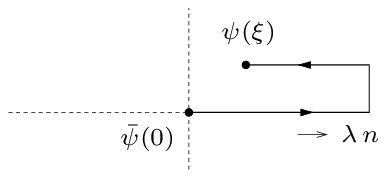}
\hspace{0.5cm}
\includegraphics[width=3.0cm]{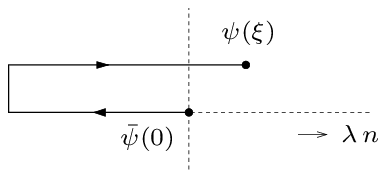}
\\
(a) $\Phi^{[+]}$
\hspace{2.4cm}
(b) $\Phi^{[-]}$
\\[0.2cm]
\includegraphics[width=3.0cm]{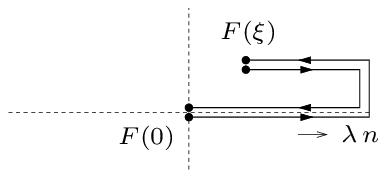}
\hspace{0.5cm}
\includegraphics[width=3.0cm]{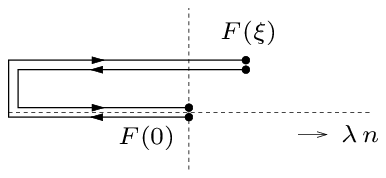}
\\
(c) $\Gamma^{[+,+]}$
\hspace{2.1cm}
(d) $\Gamma^{[-,-]}$
\\[0.2cm]
\includegraphics[width=3.0cm]{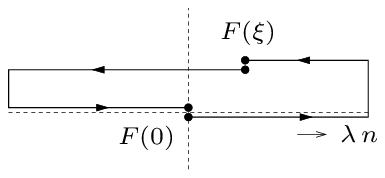}
\hspace{0.5cm}
\includegraphics[width=3.0cm]{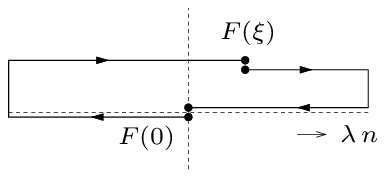}
\\
(e) $\Gamma^{[+,-]}$
\hspace{2.1cm}
(f) $\Gamma^{[-,+]}$
\caption{
Simplest structures (without loops)
for gauge links and operators in quark correlators (a)-(b)
and gluon correlators (c)-(f).
\label{simplelinks}}
\end{wrapfigure}
In the diagrammatic approach the Wilson lines arise by resumming all
collinear gluons exchanged between the soft and the hard partonic parts of the
hadronic process.
The integration path $C$ is fixed by the (color-flow structure
of) the hard partonic scattering~\cite{Bomhof:2006dp}.
Basic examples (see Fig.~\ref{simplelinks}) 
are semi-inclusive deep-inelastic scattering (SIDIS) where
for the quark correlator
the resummation of all final-state interactions leads to the future
pointing Wilson line $\mathcal U^{[+]}$,
and Drell-Yan scattering where the initial-state interactions lead to
the past  pointing Wilson line $\mathcal U^{[-]}$.
These links connect the parton fields in the correlator,
running along the light-like direction $n$, conjugate to $P$ 
(satisfying $P\cdot n = 1$ and $n^2 = 0$) and closing 
in the transverse direction at lightcone 
infinity~\cite{Belitsky:2002sm}.
For gluons the correlators including links are given by
\begin{equation}
\Gamma^{[C,C^\prime]}_{\alpha\beta}(x,p_\st;n)
=  \int \frac{d(\xi\cdot P)\,d^2\xi_\st}{(2\pi)^3}
\ e^{i\,p\cdot \xi}
\ \mbox{Tr}\left(F^{n}_{\ \ \beta}(0)\,U^{[n,C]}_{[0,\xi]}
\,F^{n}_{\ \ \alpha}(\xi)\,U^{[n,C^\prime]}_{[\xi,0]}\right)
\vert P\rangle \biggr|_{LF},
\end{equation}
with the simplest possibilities also shown in Fig.~\ref{simplelinks}.

\section{Observables}

Considering intrinsic transverse momenta is useful as it is
possible to access them in experiments. 
The collinear fractions ($x$ or $z$) in the 
Sudakov expansion of the parton momenta can be related to kinematical
ratios of hard momenta (e.g. $x \approx x_B = Q^2/2P\cdot q$ 
and $z \approx z_h = P_h\cdot P/P\cdot q$ in semi-inclusive 
deep inelastic scattering) up to $O(1/Q^2)$ corrections.
Therefore the quantity
$q_\st = q + x_B\,P - P_h/z_h \approx k_\st - p_\st$ can be
measured in semi-inclusive deep inelastic scattering (SIDIS), 
$\gamma^\ast (q) + N (P) \rightarrow h (P_h) + X$.
It is zero at leading order ($O(Q)$ in the hard scale), but relates 
to the intrinsic transverse momenta at $O(M)$.
The vector $q_\st$ is the 
transverse momentum of $q$ in a frame in which $P$ and $P_h$ are chosen 
parallel or (experimentally more useful) related to the transverse momentum
of $P_h$, $q_\st = -P_{h\perp}/z_h$ in a frame in which $q$ and $P$ are 
chosen parallel.
With $Q_\st^2 = -q_\st^2$, one needs TMD functions when $Q_\st \sim
O(M)$ and one needs a collinear description involving a subprocess with
one more parton radiated off when $Q_\st \sim O(Q)$. Matching
of these approaches was condidered in Ref.~\cite{Bacchetta:2008xw}.
Not only in electroweak processes like SIDIS or the Drell-Yan process
transverse momenta can be accessed, but one can also consider
inclusive hadron-hadron scattering. 
The experimental signature in this case is the non-collinearity
of the produced particles/jets in the plane perpendicular
to the colliding beam particles, outlined in detail in 
Ref.~\cite{Boer:2003tx}.

Accessing intrinsic transverse momenta in most cases requires a
study of azimuthal dependence in high energy processes. Although
the effects are in principle not suppressed by powers of the hard 
scale in comparison with the leading collinear treatment, it 
requires measuring hadronic scale quantities (transverse momenta)
in a high momentum environment. 
In applications to explain SSA time reversal (T) 
invariance plays an important role: 
\\
(1) The theory of QCD is T-invariant. This allows to distinguish
quantities and observables according to their T-behavior. 
Collinear correlators $\Phi(x)$ and $\Gamma(x)$, obtained after
integration over transverse momenta, are T-even. 
For the TMD correlators, however, the T-operation interchanges
$\Phi^{[+]}(x,p_\st) \leftrightarrow \Phi^{[-]}(x,p_\st)$ (and similar
relations for gluon TMD correlators), allowing  
T-even and T-odd combinations.
\\
(2) For fragmentation functions the appearance of an hadronic out-state in
the definition, prohibits
 the use of T-symmetry as a constraint and one 
has always both T-even and T-odd parts in the correlator, although one 
can separate the correlators into two classes containing 
T-even or T-odd operator combinations in analogy with the case 
of distributions, referred to as naive T-even or T-odd.
\\
(3) In a scattering process, in which T-symmetry can be used as a constraint,
SSA would be forbidden. In fact the only real 
example of this is DIS (omitting electromagnetic interaction
effects). For hadron-hadron scattering, e.g.\ the Drell-Yan process, one 
has a two-hadron initial state and only the assumption of a factorized
description would imply absence of SSA.
We now know that this assumption is not valid, even not at leading 
order!
Similarly for processes with identified hadrons in the final state 
T-invariance does not give constraints.
\\
(4) At leading order in $\alpha_s$, however, it is possible to connect
SSA (being T-odd observables) to the 
T-odd soft parts, since the hard process
will be T-even at this leading order. Collins and Sivers effects as
explanation for SSA are the best known examples.

\section{TMD treatment}

As already referred to in section~\ref{TMDcorrelators} the
gauge links in the correlators are the result of resumming
leading matrix elements with collinear gluons. 
The presence of links, differing for each 
partonic sub-diagram and its color-flow,
results in the following expression for a hard cross section 
at measured $q_\st$ (involving in general complex diagram-dependent
gauge-link paths),
\be
\frac{d\sigma}{d^2q_\st}\sim \sum_{D,abc\ldots}
\Phi_a^{[C_1(D)]}(x_1,p_{1\st})\,\Phi_b^{[C_2(D)]}(x_2,p_{2\st})
\,\hat\sigma_{ab\rightarrow c\ldots}^{[D]}
\Delta_c^{C_1^\prime(D)]}(z_1,k_{1\st})\ldots + \ldots
\label{basic}
\ee
where the sum $D$ runs over diagrams distinguishing also the
color flow and $abc\ldots$ is the summation over quark
and antiquark flavors and gluons. Dirac and Lorentz indices,
traces are suppressed. The ellipsis at the end indicate
contributions of other hard processes.

The results for cross sections after integration over the transverse
momenta $q_\st$ involve the path-independent integrated correlators
$\Phi(x)$ rather than
the path-dependent TMD correlators $\Phi^{[C(D)]}(x,p_\st)$. Thus, from
Eq.~\ref{basic} one gets the well-known result
\be
\sigma \sim \sum_{abc\ldots} \Phi_a(x_1)\,\Phi_b(x_2)
\,\hat\sigma_{ab\rightarrow c\ldots}\Delta_c(z_1)\ldots + \ldots,
\ee
where
$\hat \sigma_{ab\rightarrow c\ldots}
= \sum_{D} \hat\sigma^{[D]}_{ab\rightarrow c\ldots}$
is the partonic cross section.

Constructing a weighted cross section (azimuthal asymmetry) by
including a weight $q_\st^\alpha$ in the $q_\st$-integration leads
to the {\em transverse moments}
\be
\Phi_\partial^{\alpha\,[C]}(x) =
\int d^2p_\st\ p_\st^\alpha\Phi^{[C]}(x,p_\st)
=\widetilde\Phi_\partial^{\alpha}(x)
+ C_G^{[U(C)]}\,\pi\Phi_G^\alpha(x,x).
\label{decomposition}
\ee
These moments still contain a path dependence, so Eq.~\ref{basic} cannot be
simplified immediately but as shown the path dependence is contained in 
a (gluonic pole) factor $C_G$, which can easily be calculated. The first term,
$\widetilde\Phi_\partial(x)$, is a collinear correlator
containing matrix elements with T-even operators, while $\Phi_G(x,x-x_1)$ is
a collinear correlator with a structure of a quark-gluon-quark
correlator involving the gluon field $F^{n\alpha}$.
In Eq.~\ref{decomposition} one needs the zero-momentum ($x_1 = 0$) limit
for the gluon momentum. This matrix element is known as the gluonic pole
matrix element. The operators involved are T-odd. Both collinear
correlators on the RHS in Eq.~\ref{decomposition} are link-independent.
Using this decomposition one can write down a 
parton-model like expansion for the single-weighted cross section
$\left< q_\st^\alpha\sigma\right>$ in which 
$\widetilde\Phi_{\partial}^{\alpha}(x)$ is multiplied with the partonic
cross section, while $\pi\Phi_{G}^{\alpha}(x,x)$ is multiplied
with the {\em gluonic pole cross section},
$\hat \sigma_{[a]b\rightarrow c\ldots}
= \sum_{D} C_G^{[U(C(D))]}\hat\sigma^{[D]}_{ab\rightarrow c\ldots}$, 
which just like the normal partonic cross sections also
constitutes a different gauge-invariant combination of the squared 
amplitudes~\cite{Bomhof:2006ra}. For more complex weightings or
trying to stay at the unintegrated level, one has to make additional
assumptions outlined in Ref.~\cite{Bomhof:2007xt}.
In this paper also the split-up of TMD functions in 
\be
\Phi^{[U]}(x,p_\st) = 
\tfrac{1}{2}\left(\Phi^{[{\rm even}]}(x,p_\st) + 
G_G^{[U]}\,\Phi^{[{\rm odd}]}(x,p_\st)\right) + 
\delta\Phi^{[U]}(x,p_\st) ,
\ee
with $\Phi^{[{\rm even/odd}]} = \tfrac{1}{2}(\Phi^{[+]} \pm \Phi^{[-]})$
is discussed, with $\delta\Phi^{[U]}(x,p_\st)$ satisfying
$\delta\Phi^{[U]}(x) =  \delta\Phi_\partial^{\alpha[U]}(x) =  0$.

The approach to understand T-odd observables like single spin asymmetries
via the TMD correlators and the non-trivial gauge link structure unifies
a number of approaches to understand such observables, in particular
the collinear approach and the inclusion of soft gluon interactions.
Although the treatment of fragmentation correlators also separates into
parts with T-even and T-odd operator structure, gluonic pole
contributions (T-odd parts) in the case of fragmentation might vanish.
Indications come from the soft-gluon approach~\cite{Collins:2004nx}
and a recent spectral analysis in a spectator model
approach~\cite{Gamberg:2008yt}.


\begin{footnotesize}


\end{footnotesize}


\end{document}